\documentclass[5p,sort&compress]{elsarticle}
\usepackage{amssymb}
\usepackage{amsmath}
\usepackage{graphicx}
\usepackage{hyperref}
\usepackage[utf8]{inputenc}
\usepackage{color}

\newcommand{\be}{\begin{equation}}
\newcommand{\ee}{\end{equation}}
\newcommand{\ba}{\begin{eqnarray}}
\newcommand{\ea}{\end{eqnarray}}

\begin{document}

\title{Fermion number violating effects in low scale leptogenesis}

\author[epfl]{Shintaro Eijima}
\ead{Shintaro.Eijima@epfl.ch}
\author[epfl]{Mikhail Shaposhnikov}
\ead{Mikhail.Shaposhnikov@epfl.ch}

\address[epfl]{
 Institute of Physics, Laboratory for Particle Physics and Cosmology,
\'{E}cole Polytechnique F\'{e}d\'{e}rale de Lausanne, CH-1015 Lausanne, 
Switzerland}


\begin{abstract}
The existence of baryon asymmetry and dark matter in the Universe may be related to CP-violating reactions of three heavy neutral leptons (HNLs) with masses well below the Fermi scale.  The dynamical description of the lepton asymmetry generation, which is the key ingredient of baryogenesis and of dark matter production, is quite complicated due to the presence of many different relaxation time scales and the necessity to include quantum-mechanical coherent effects  in HNL oscillations. We derive  kinetic equations accounting for fermion number violating  effects missed so far and identify one of the domains of HNL masses that can potentially lead to large lepton asymmetry generation boosting the sterile neutrino dark matter production.

\end{abstract}

\maketitle

\section{Introduction}
\label{intr}
Though the canonical Standard Model (SM)  has been completed by the discovery of the Higgs boson and may be a valid effective quantum field theory all the way up to the Planck scale  (for recent discussions see \cite{Bezrukov:2012sa,Buttazzo:2013uya,Bednyakov:2015sca})   it is inconsistent with a number of observations. They include the non-zero neutrino masses, the presence of Dark Matter (DM) in the Universe, and its baryon asymmetry (BAU). Perhaps, the most minimal way to address all these problems on the same footing is to extend the SM by three right-handed neutrinos  with masses below the Fermi scale \cite{Asaka:2005an,Asaka:2005pn}. These new fermions $N_I$, I=1,2,3 (following the Particle Data Group \cite{Agashe:2014kda} we will call them Heavy Neutral Leptons or HNLs for short) are singlets with respect to the SM gauge group and thus are allowed to have Majorana neutrino masses. The lightest of these particles, $N_1$,  may play a role of Dark Matter \cite{Dodelson:1993je,Shi:1998km}. Two others ($N_2$ and $N_3$), if (almost) degenerate, can produce the baryon asymmetry of the Universe \cite{Akhmedov:1998qx},  \cite{Asaka:2005pn}  and explain non-zero neutrino masses and mixings at the same time. This model was dubbed the $\nu$MSM for ``Neutrino Minimal Standard Model'' \cite{Asaka:2005an}. For a number of computations of baryon asymmetry in this model see \cite{Shaposhnikov:2006nn,Shaposhnikov:2008pf,Canetti:2010aw,Asaka:2010kk,Asaka:2011wq,Canetti:2012vf,Canetti:2012kh,Shuve:2014zua,Abada:2015rta,Hernandez:2016kel,Drewes:2016gmt}.

The most conservative scenario of the Universe evolution, which does not require any new physics beyond the $\nu$MSM, proceeds as follows. First, the Universe is inflated by the SM Higgs field \cite{Bezrukov:2007ep} and heated up due to Higgs field oscillations to temperatures $T\sim 10^{14}$ GeV \cite{Bezrukov:2008ut,GarciaBellido:2008ab,Bezrukov:2014ipa}. The Higgs inflation prepares the initial conditions for the Hot Big Bang \cite{Bezrukov:2008ut} at $T\sim 10^{14}$ GeV: baryon and lepton numbers of the Universe are equal to zero, and the number densities of HNLs at this time  are zero as well. The particles $N_2$ and $N_3$ enter into thermal equilibrium  below the sphaleron freeze-out temperature $T_{sph} \simeq 130$ GeV and produce  baryon asymmetry of the Universe  in a set of  processes which include their coherent oscillations, transfer of lepton number from HNLs to active leptons and back \cite{Akhmedov:1998qx}, \cite{Asaka:2005pn}, and rapid anomalous sphaleron transitions \cite{Kuzmin:1985mm}. The lighter HNL -- $N_1$ -- DM sterile neutrino never equilibrates and is mainly produced at temperatures $T_{DM} \sim 100-300$ MeV by transitions from the ordinary neutrinos to $N_1$ \cite{Dodelson:1993je,Shi:1998km,Abazajian:2001nj,Asaka:2006rw,Asaka:2006nq,Laine:2008pg,Ghiglieri:2015jua}.  The combination of X-ray and Lyman-$\alpha$ bounds on the DM sterile neutrino excludes  the ``non-resonant'' Dodelson-Widrow mechanism \cite{Dodelson:1993je} for their production, which operates in the cosmic plasma with small lepton asymmetries. In other words, to get enough DM particles $N_1$, the processes involving $N_{2,3}$ should produce \cite{Shaposhnikov:2008pf} sufficiently large lepton asymmetry $\Delta L/L > 2 \times 10^{-3}$ which must be present at temperatures $T_{DM}$.  This is needed to boost the production of $N_1$ due to the resonant mechanism proposed by Shi and Fuller in \cite{Shi:1998km} and developed in a rigorous way in \cite{Asaka:2006rw,Asaka:2006nq,Laine:2008pg,Ghiglieri:2015jua}. The production of this large lepton asymmetry must take place below the sphaleron temperature $T_{sph}$, otherwise the baryon asymmetry will be too large \cite{Shaposhnikov:2008pf}.

The estimates of the equilibration rates of $N_{2,3}$ in \cite{Shaposhnikov:2008pf,Canetti:2012vf,Canetti:2012kh} and in more recent works \cite{Ghisoiu:2014ena,Ghiglieri:2016xye} based on careful thermal field theory computations showed that for all parameter choices consistent with observed pattern of neutrino masses and oscillations the HNLs $N_{2,3}$ enter in thermal equilibrium at some temperature $T_\text{in}$ exceeding tens of GeV and go out of thermal equilibrium at temperatures $T_\text{out}<T_\text{in}$ which can be as small as $1$ GeV. This has led to the conclusion that the equilibrium period between $T_\text{in}$ and $T_\text{out}$ erases all the lepton asymmetry which could have been generated at freeze-in temperature $T_\text{in}$ ,  requiring that the large lepton asymmetry needed for effective dark matter production must be created at $T<T_\text{out}$ .  The analysis made in \cite{Canetti:2012kh} demonstrated that a large lepton asymmetry can indeed be generated in the scattering processes involving $N_{2,3}$ at the freeze-out temperature $T_\text{out}$ and below it in out-of-equilibrium decays of $N_{2,3}$. This asymmetry does not exceed $\Delta L/L \simeq 3\times 10^{-2}$, leading to the conclusion that the mass of the DM sterile neutrino must lie in the interval from $1$ to $50$ keV, to be consistent with the Lyman-$\alpha$ and  phase density constraints coming from observations of dwarf galaxies \cite{Boyarsky:2008ju,Gorbunov:2008ka}. As for $N_{2,3}$, their {\em physical masses} should be between $1.5$ GeV and  $\simeq 80$ GeV (the $W$-boson mass) and be extremely degenerate, $\Delta M_{phys}/M < 10^{-15}$ \cite{Shaposhnikov:2008pf,Roy:2010xq,Canetti:2012vf,Canetti:2012kh,Blondel:2014bra}\footnote{These restrictions are not required if the DM sterile neutrino is produced by some new interactions not contained in the renormalizable $\nu$MSM Lagrangian \cite{Shaposhnikov:2006xi,Kusenko:2006rh,Bezrukov:2011sz}.} . The latter condition comes from the requirement that the period of $N_2 \leftrightarrow N_3$ oscillations should be comparable with the age of the Universe at the time of lepton asymmetry production, to insure the resonance \cite{Shaposhnikov:2008pf,Roy:2010xq}. The minimal scenario that has been proven to work, albeit under the requirement of a {\em strong fine-tuning} (of the order of $10^{-4}$ \cite{Canetti:2012kh}) between two different contributions to the physical mass difference: one coming from the Yukawa couplings and the Higgs condensate, and another from Majorana masses of $N_{2,3}$. 

The aim of the present paper is to show that the part of the lepton asymmetry generated at $T_\text{in}$ {\em can in fact survive until the temperatures of sterile neutrino DM production} $\sim 100$ MeV, in-spite of the fact that HNLs are well in thermal equilibrium between $T_\text{in}$ and $T_\text{out}$. Qualitatively, this comes about because of the following reasons. In the symmetric phase of the electroweak theory the transfer of asymmetry from active to sterile sector and back occurs mainly via the processes with fermion number conservation (we attribute positive fermion number to left-handed neutrinos and to right-handed HNLs) with the rate $\Gamma_+$.  The rate of fermion number non-conserving processes $\Gamma_-$ is suppressed by a kinematic factor $(M/k)^2$, where $M$ is the HNL mass, and $k\sim 3T$ is the typical momentum of fermions in the plasma. 

On the contrary, in the Higgs phase, at temperatures of the order of tens GeV, the dominant reaction  is induced by the mixing term between $\nu$'s and $N$'s and has a rate $\Gamma_-$ exceeding that of the Universe expansion at  $T_\text{out}<T<T_\text{in}$. It proceeds with fermion number non-conservation:
 left handed neutrinos go into left-handed anti-HNLs and vice-versa. 

In a large portion of the $\nu$MSM parameters the reactions with fermion number  conservations are faster and give the main contribution to baryogenesis at the sphaleron freeze-out temperature $T\simeq 130$ GeV. However,  in a specific domain of the NHL masses and couplings, the rate of these processes never exceeds  the Hubble rate. Thus, the asymmetry in this almost conserved number is {\em protected from dilution}, in-spite of the fact that HNLs are equilibrated due to  the processes with fermion number violation\footnote{The importance of the processes with and without fermion number violation has been already realised in \cite{Shaposhnikov:2008pf}. Unfortunately, the estimates of the relevant rates were not done correctly and the conclusions we arrived in our present work were not achieved at that time. Some further studies of the processes with fermion number non-conservation in this context were carried out in \cite{Hambye:2016sby}.}. Moreover, at $T \simeq T_\text{in}$ the rate $\Gamma_+$ can be close to the Hubble rate, meaning that  large  asymmetry in this number {\em can be generated}. To understand whether it is indeed produced would require numerical solution of our integro-differential kinetic equations for many parameters of the  $\nu$MSM,  which is not attempted here.

The paper is organised as follows. In Section \ref{kin} we will derive kinetic equations accounting for helicity structure of HNL interactions. In Section \ref{rates} we analyse the different rates and identify the range of $\nu$MSM parameters which may potentially lead to large lepton asymmetries surviving  until small temperatures where the production of DM sterile neutrino takes place. In Section \ref{concl} we summarise our results.

\section{Kinetic equations and helicity}
\label{kin}
In this section we derive the kinetic equations describing the evolution of HNLs density matrix and lepton densities.  To elucidate their structure, we will consider first the temperatures well below the electroweak scale, deeply in the Higgs phase. We also neglect for the time being the subtleties related to electric neutrality of the plasma \cite{Khlebnikov:1996vj,Laine:1999wv,Nardi:2006fx,Bodeker:2014hqa} and consider the system with HNLs and active neutrinos only, this will be corrected towards the end of this Section. We take a pair of almost degenerate HNLs $N_2$ and $N_3$, the generalisation to the case of arbitrary number of $N$ is straightforward. It is convenient to unify $N_2$ and $N_3$ in one Dirac spinor, as has been done in \cite{Shaposhnikov:2008pf}, and consider the $\nu$MSM Lagrangian 
\begin{align}
 \mathcal{L} &=  \mathcal{L}_{SM} +   \overline{\Psi} i \partial_\mu \gamma^\mu \Psi
 - M \overline{\Psi}\Psi +  \mathcal{L}_{int}, \nonumber\\
\mathcal{L}_{int} &=
- \frac{\Delta M}{2} (\overline{\Psi}\Psi^c + \overline{\Psi^c}\Psi) \nonumber \\
 &- (h_{\alpha 2} \langle \Phi \rangle \overline{\nu_{L \alpha}} \Psi + h_{\alpha 3} \langle \Phi \rangle \overline{\nu_{L\alpha}} \Psi^c + h.c.),
 \label{eq:L}
\end{align}
where $\mathcal{L}_{SM}$ is the SM part, $\Psi = N_2 + N_3^c$ is the HNL field in the pseudo-Dirac basis, $M = (M_{3} + M_{2})/2$ and $\Delta M = (M_{3} - M_{2})/2$ are the common mass and Majorana mass difference of HNLs, respectively, $h_{\alpha I}$ is a matrix of Yukawa coupling constants and $\langle \Phi \rangle$ is the temperature dependent Higgs vacuum expectation value\footnote{At small temperatures we are working now the dynamical character of the Higgs field is not important.}, which is $174.1$~GeV at zero temperature.
The HNL field $\Psi$ is given in terms of creation and annihilation operators by
\begin{align}
 \Psi &= \int \frac{d^{3}k}{(2 \pi)^{3}}\frac{1}{\sqrt{2 k_{0}}} \sum_{\sigma} \left[ a_{\sigma}(k) u_{\sigma}(k) e^{- i k \cdot x} \right. \nonumber \\
        & \left. \hspace{90pt}+ b^{\dagger}_{\sigma}(k) v_{\sigma}(k) e^{i k \cdot x} \right], 
\end{align}
where $\sigma=\pm$ describes the HNL helicity. The attribution of fermion numbers to $a_{\sigma},~ b_{\sigma}$ leading to fermion number conservation in the limit $M\to 0,~\Delta M \to 0$ is shown in Table~\ref{tab:ops}.
\begin{table}[!tb]
 \centering
 \begin{tabular}{|c|c|} \hline
   plus, particles & minus, anti-particles \\ \hline
  $a^{\dagger}_{+}(k), \, b^{\dagger}_{-}(k)$ & $a^{\dagger}_{-}(k), \, b^{\dagger}_{+}(k)$ \\ \hline
 \end{tabular}
 \caption{Fermion numbers of creation operators.}
 \label{tab:ops}
\end{table}

We will consider $\mathcal{L}_{int}$ as a perturbation and work in the second order in Yukawa couplings and first order in $\Delta M$ assuming for power counting that $M\Delta M \sim h^2\langle \Phi \rangle^2$, where $h$ is a typical value of the Yukawa couplings. The quadratic mixing term $\mathcal{L}_{int}$ leads to communication between sterile sector of HNLs and the rest of the SM, ensuring creation and destruction of HNLs, their coherent oscillations, lepton number non-conservation, and transfer of asymmetries from active flavours to sterile and back.

To construct kinetic equations, we work in the Heisenberg picture of quantum mechanics and use the ideas of  \cite{Sigl:1992fn} in what follows\footnote{The novel feature of our derivation is that we keep the evolution of rapidly oscillating products of creation and annihilation operators of the type $a^\dagger b^\dagger,~  a^\dagger a^\dagger,~ab$, etc, playing a crucial role for description of the processes with fermion number conservation.}. 
The derivation is performed by using four creation operators  $a^\dagger_\sigma(k), b^\dagger_\sigma(k)$ for HNLs and two (for each generation $\alpha$) neutrino and antineutrino operators, $a^\dagger_{\nu_\alpha}(k), b^\dagger_{\nu_\alpha}(k)$. 

Let $\rho$ be a (time-independent) density matrix of the complete system. The HNL abundances, including coherent quantum-mechanical correlations between $N_2$ and $N_3$, are given by the averages $\text{Tr} [a^\dagger_\sigma(k)a_{\sigma'}(k)\rho]$, $\text{Tr} [b^\dagger_\sigma(k)b_{\sigma'}(k)\rho]$, $\text{Tr} [a^\dagger_\sigma(k)b_{\sigma'}(k)\rho]$,  and  $\text{Tr} [b^\dagger_\sigma(k)a_{\sigma'}(k)\rho]$. The neutrino number densities are $a^\dagger_{\nu_\alpha}(k)a_{\nu_\beta}(k)$ and $b^\dagger_{\nu_\alpha}(k)b_{\nu_\beta}(k)$. Symbolically, all these number-density operators will be denoted by $Q^0_i$. With the use of our generic notation $Q^0_i$, the abundances are given by $q^0_i= \text{Tr}[Q^0_i\rho]$. 

The time derivative of $q^0_i$ is readily found:
\begin{align}
 i \dot{q^0_i} = \text{Tr}\left([{\bf H},Q^0_i]\rho\right)~,
 \label{eq:FP}
\end{align}
where ${\bf H}$ is the total Hamiltonian of the system under consideration. Since we neglected the existence of charged leptons for the moment, the commutator of $Q^0_i$ with the SM Hamiltonian is zero, the only non-trivial contribution comes from the interaction Hamiltonian $H_{int}$, associated with $\mathcal{L}_{int}$ defined in  (\ref{eq:L}). It is easy to see that the commutators $[H_{int},Q^0_i]$  are the quadratic polynomials with respect to creation and annihilation operators, containing all possible terms, to list just a few:  $a^\dagger_\sigma(k) b_{\nu_\beta}(k)$, $a^\dagger_\sigma(k) b^\dagger_{\nu_\beta}(k)$,  $a_\sigma(k) b_{\nu_\beta}(-k)$. These operators are multiplied by the first power of Yukawa couplings or by $\Delta M$. Denote by  $Q^1_i$ these binomials, and by $q^1_i=\text{Tr}[Q^1_i\rho]$.  Now we continue the chain, and write an equation for every $q^1_i$ which appeared at the first step:
\begin{align}
 i \dot{q^1_i} = \text{Tr}\left([{\bf H},Q^1_i]\rho\right)~,
 \label{eq:FP1}
\end{align}
and then repeat the procedure again and again. In this way we get an infinite chain of kinetic equations, which includes the averages of higher and higher polynomials in creation and annihilation operators. 

To truncate the system, we proceed as follows. The total Hamiltonian of the system can be written as
\begin{align}
{\bf H} = H_2+H_{int}+H_{int}^{SM},
 \label{eq:ham}
\end{align}
where $H_2$ is the quadratic part including HNLs and active neutrinos, and  $H_{int}^{SM}$ describes the SM interactions. As an example, let us take an operator $Q^1_1=a^\dagger_\sigma(k) b_{\nu_\beta}(k)$. Its commutator with the total Hamiltonian contains 3 terms. The first one is 
\begin{align}
[Q_1^1,H_2]= - (E_N(k) - \epsilon_{\nu_{\beta}} (k)) Q_1^1, 
 \label{eq:h0}
\end{align}
where $E_N(k)=\sqrt{k^2+M^2}$ and $\epsilon_\nu(k)=k$ are the energies of HNL and active neutrino respectively (the small neutrino mass can be safely neglected here). The second is $[Q_1^1,H_{int}]= \Sigma_i C_i Q^0_i$ where $C_i$ are the coefficients containing Yukawa couplings and $\Delta M$. The third one is $[Q_1^1,H_{int}^{SM}] $ and is of the order of Fermi constant $G_F$ and of 4th order in creation and annihilation operators. It accounts for neutrino interaction in the medium. The first two terms contain the operators that have already showed up in the first kinetic equation  (\ref{eq:FP}), while the third one contains new operators. To find their time evolution  would require next steps in the iterative procedure. 

To deal with the third term, we note that the neutrino interactions in the plasma can be accounted for by modification of neutrino energy $\epsilon_\nu(k)$,  replacing it with the temperature dependent dispersion relation $E_\nu(k)$ (related to the real part of neutrino propagator $\Sigma$) and by attributing to it an imaginary part $\gamma_\nu(k)>0$ (associated with absorptive part of $\Sigma$). These considerations suggest the following modification of the commutation relations:
\begin{align}
[a_{\nu_\alpha}(k),H_2] \to (E_\nu(k)+i\gamma_\nu (k)/2) a_{\nu_\alpha}(k), \\
[a^\dagger_{\nu_\alpha}(k),H_2] \to -(E_\nu(k)-i\gamma_\nu (k)/2) a^\dagger_{\nu_\alpha}(k),
\label{eq:hnew}
\end{align}
where the signs in front  of $\gamma_\nu(k)$ are chosen in such a way that they correspond to damping rather than an instability. The similar rules apply to antineutrinos. These substitutions effectively account for the third term which can now be removed. 

The system of kinetic equations for  $q^0_i$ and $q^1_i$ is now complete. It can be simplified even further as the active neutrinos are well in thermal equilibrium at all  temperatures we are interested it, $\gamma_\nu/H \gg 1$, where $H$ is the Hubble rate. Again, we take $Q_1^1$ as an example. The equation for it now reads
\begin{align}
 i \dot{q^1_1} = - \left(E_N(k)-(E_\nu(k)+i\gamma_\nu/2)\right) q_1^1 + \Sigma_i C_i q^0_i~,
 \label{eq:11}
\end{align}
and has an approximate slow varying solution 
\begin{align}
 q^1_1 = \frac{ \Sigma_i C_i q^0_i}{\left(E_N(k)-(E_\nu(k)+i\gamma_\nu/2)\right)}.
 \label{eq:q11}
\end{align}
All $q^1_i$ can be found in this way and inserted into equation  (\ref{eq:FP}) for $q^0_i$. As a result, we get the kinetic description in terms of $q^0_i$ only.

The realisation of this program requires a straightforward but tedious computation, which we have done with the use of DiracQ Mathematica package \cite{Wright:2013sv}.  By introducing the notations
\begin{align}
\rho_{\nu_\alpha}&=\text{Tr}[a^{\dagger}_{\nu_{\alpha}}(k) a_{\nu_{\alpha}}(k) \rho] - \rho_{\nu}^{eq}, 
\label{eq:notat_nu} \\
\rho_{\bar{\nu}_\alpha}&=\text{Tr}[b^{\dagger}_{\nu_{\alpha}}(k) b_{\nu_{\alpha}} (k) \rho] - \rho_{\nu}^{eq}, 
\label{eq:notat_bnu} 
\end{align}
\begin{align}
\rho_{N}&=\begin{pmatrix}
  \text{Tr}[a^{\dagger}_{+}(k) a_{+}(k) \rho] & \text{Tr}[a^{\dagger}_{+}(k) b_{-}(k)\rho] \\
  \text{Tr}[b^{\dagger}_{-}(k) a_{+}(k) \rho] & \text{Tr}[b^{\dagger}_{-}(k) b_{-}(k) \rho]
\end{pmatrix} - \rho_{N}^{eq} \mathbf{1}, 
\label{eq:notat_N} 
\end{align}
\begin{align}
\rho_{\bar{N}}&=\begin{pmatrix}
  \text{Tr}[a^{\dagger}_{-}(k) a_{-}(k) \rho] & \text{Tr}[a^{\dagger}_{-}(k) b_{+}(k) \rho] \\
  \text{Tr}[b^{\dagger}_{+}(k) a_{-}(k) \rho] & \text{Tr}[b^{\dagger}_{+}(k) b_{+}(k) \rho]
\end{pmatrix} - \rho_{N}^{eq} \mathbf{1},
\label{eq:notat_bN}
\end{align}
where $\rho_{\nu}^{eq}$ and $\rho_{N}^{eq}$ are equilibrium distribution functions of neutrinos and HNLs, and ``$\mathbf{1}$'' is the unity matrix, we arrived to the following result\footnote{We do not account here the expansion of the universe, but it can be easily accommodated.}
\begin{align}
i \, \frac{d\rho_{\nu_\alpha}}{dt} 
&= - i \, \Gamma_{\nu_\alpha} \rho_{\nu_\alpha}
    + i \, \text{\text{Tr}}[\tilde{\Gamma}_{\nu_\alpha} \, \rho_{\bar{N}}], 
\label{eq:KEnu}\\
i \, \frac{d\rho_{\bar{\nu}_\alpha}}{dt} 
&= - i \, \Gamma_{\nu_\alpha}^\ast \rho_{\bar{\nu}_\alpha}
    + i \, \text{\text{Tr}}[\tilde{\Gamma}_{\nu_\alpha}^\ast \, \rho_{N}], 
\label{eq:KEnubar}\\
i \, \frac{d\rho_{N}}{dt} 
&= [H_N, \rho_N]
    - \frac{i}{2} \, \{ \Gamma_{N} , \rho_{N} \}
    + i \, \sum_\alpha \tilde{\Gamma}_{N}^\alpha \rho_{\bar{\nu}_\alpha},  
\label{eq:KEN}\\
i \, \frac{d\rho_{\bar{N}}}{dt} 
&= [H_N^\ast, \rho_{\bar{N}}]
    - \frac{i}{2} \, \{ \Gamma_{N}^\ast , \rho_{\bar{N}} \}
    + i \, \sum_\alpha (\tilde{\Gamma}_{N}^\alpha)^\ast \rho_{\nu_\alpha} .
\label{eq:KENbar}
\end{align}
The effective Hamiltonian is  
\begin{align}
 H_N &= H_0 + H_I, \\
 H_0 &= -\frac{\Delta M M}{E_N} \sigma_{1} \\
 H_I &= h_{+} \sum_{\alpha} Y_{+, \alpha}^{N} + h_{-} \sum_{\alpha} Y_{-, \alpha}^{N},
\end{align}
where $\sigma_1$ is the Pauli matrix. The production rates for HNLs are  
\begin{align}
\Gamma_N &= \Gamma_+ + \Gamma_-,\\
\label{ga+}\Gamma_+ &= \gamma_+ \sum_{\alpha} Y_{+, \alpha}^{N}, \\
\label{ga-}\Gamma_- &= \gamma_- \sum_{\alpha} Y_{-, \alpha}^{N}, \\
 \tilde{\Gamma}_N^\alpha &= - \gamma_{+} Y_{+, \alpha}^{N} + \gamma_{-} Y_{-, \alpha}^{N}, 
\end{align}
and those for active neutrinos are 
\begin{align}
 \Gamma_{\nu_\alpha} &= (\gamma_+ + \gamma_-) \sum_I h_{\alpha I} h_{\alpha I}^\ast, \\
 \tilde{\Gamma}_{\nu_\alpha} &= - \gamma_{+, \alpha}^{\nu} Y_{+,\alpha}^{\nu}+ \gamma_{-, \alpha}^{\nu} Y_{-,\alpha}^{\nu}.
\end{align} 
The coefficients are given by 
\begin{align}
\label{hplus}
 h_+ &= \frac{2 \langle \Phi \rangle^2 E_\nu (E_N + k) (E_N + E_\nu)}
            {k E_N (4(E_N+E_\nu)^2 +\gamma_\nu^2)}, \\
 h_- &= \frac{2 \langle \Phi \rangle^2 E_\nu (E_N - k) (E_N - E_\nu)}
            {k E_N (4(E_N-E_\nu)^2 +\gamma_\nu^2)}, \\
 \gamma_+ &= \frac{2 \langle \Phi \rangle^2 E_\nu (E_N + k) \gamma_\nu}
            {k E_N (4(E_N+E_\nu)^2 +\gamma_\nu^2)}, 
            \label{eq:gPb}\\
 \gamma_- &= \frac{2 \langle \Phi \rangle^2 E_\nu (E_N - k) \gamma_\nu}
            {k E_N (4(E_N-E_\nu)^2 +\gamma_\nu^2)}, 
            \label{eq:gMb}
\end{align}
and the matrices of Yukawa coupling constants are 
\begin{align}
\label{yplus}
Y_{+, \alpha}^{N} &=
\begin{pmatrix}
  h_{\alpha 3} h_{\alpha 3}^\ast & - h_{\alpha 3} h_{\alpha 2}^\ast \\
 - h_{\alpha 2} h_{\alpha 3}^\ast & h_{\alpha 2} h_{\alpha 2}^\ast
\end{pmatrix}, \\
\label{yminus}
Y_{-, \alpha}^{N} &=
\begin{pmatrix}
  h_{\alpha 2} h_{\alpha 2}^\ast & - h_{\alpha 3} h_{\alpha 2}^\ast \\
 - h_{\alpha 2} h_{\alpha 3}^\ast & h_{\alpha 3} h_{\alpha 3}^\ast
\end{pmatrix}, \\
Y_{+, \alpha}^{\nu} &=
\begin{pmatrix}
  h_{\alpha 3} h_{\alpha 3}^\ast & - h_{\alpha 2} h_{\alpha 3}^\ast \\
 - h_{\alpha 3} h_{\alpha 2}^\ast & h_{\alpha 2} h_{\alpha 2}^\ast
\end{pmatrix}, \\
Y_{-, \alpha}^{\nu} &=
\begin{pmatrix}
  h_{\alpha 2} h_{\alpha 2}^\ast & - h_{\alpha 2} h_{\alpha 3}^\ast \\
 - h_{\alpha 3} h_{\alpha 2}^\ast & h_{\alpha 3} h_{\alpha 3}^\ast
\end{pmatrix},
\end{align}
where  $E_\nu=k-b_L$ and function $b_L$ is often called neutrino potential in the medium. It has been computed in a number of papers in different limits \cite{Notzold:1987ik,Morales:1999ia}. The neutrino damping rate as well as $b_L$ can be taken from a recent work \cite{Ghiglieri:2016xye}. 

\begin{figure}[!tb]
  \begin{center} 
    \includegraphics[clip, width=8cm]{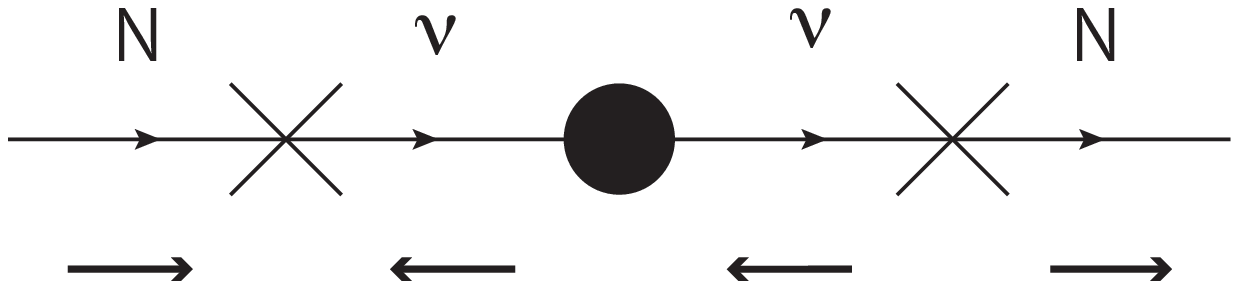} \\
    \vspace{10pt}
    (a) Fermion number conserving process \\
    \vspace{-75pt}
    \includegraphics[clip, width=8cm]{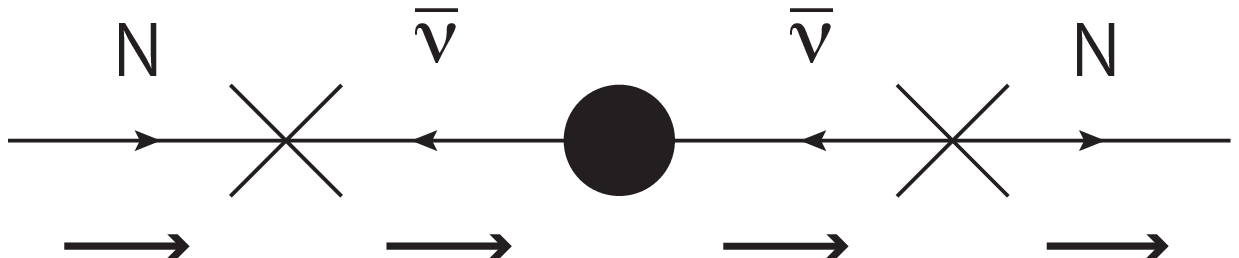} \\
    \vspace{10pt}
    (b) Fermion number violating process 
    \caption{Diagrammatical descriptions of fermion number conserving and violating processes. The arrows under particle lines show the direction of momentum, and the black spot indicates interactions in plasma.}
    \label{fig:diagram}
  \end{center}
\end{figure}
The parts of the Hamiltonian and production rates with subscript ``$+$'' and ``$-$''  are associated with the fermion number conserving and violating operators, respectively. In Figure~\ref{fig:diagram} they are expressed diagrammatically, where the vertexes (denoted by the cross) in (a) come from the structures in $H_{int}$ containing the product of two creation (or annihilation) operators, as for example in (\ref{square})
\begin{align}
\begin{split}
h_{\alpha 2} \ a_{+}(k) \ b_{\nu}(-k) \ e^{- i (E_{N}+E_{\nu})t}, \\
h_{\alpha 2}^{\ast} \ b_{\nu}^{\dagger}(-k) \ a_{+}^{\dagger}(k) \ e^{i (E_{N}+E_{\nu})t}, 
\label{square}
\end{split}
\end{align}
and those in (b) coming from a product of creation and annihilation operators as, for instance:
\begin{align}
\begin{split}
h_{\alpha 3}^{\ast} \ a_{+}(k) \ b_{\nu}(k)^{\dagger} \ e^{- i (E_{N}-E_{\nu})t}, \\
h_{\alpha 3} \ b_{\nu}(k) \ a_{+}^{\dagger}(k) \ e^{i (E_{N}-E_{\nu})t}.
\end{split}
\end{align}

The structure of the kinetic equations is exactly the same as it was first found in \cite{Asaka:2005pn} and elucidated in \cite{Shaposhnikov:2008pf}\footnote{The kinetic equations derived in the seminal work \cite{Akhmedov:1998qx} unfortunately are not correct as they do not contain the transfer (last) terms in Eqs.~(\ref{eq:KEN},\ref{eq:KENbar}) and the equations for lepton asymmetries (\ref{eq:KEnu},\ref{eq:KEnubar}).}, but now with all the terms expressed explicitly through parameters of the theory. The same set of equations (\ref{eq:KEnu}-\ref{eq:KENbar}) was used in  \cite{Shaposhnikov:2006nn,Canetti:2010aw,Asaka:2010kk,Asaka:2011wq,Canetti:2012vf,Canetti:2012kh,Shuve:2014zua,Abada:2015rta,Hernandez:2016kel,Drewes:2016gmt}, for analysis of baryon asymmetry generation in the $\nu$MSM, but with different choices of kinetic coefficients. The novel results of our work are the formulas for the Hamiltonian and different rates (\ref{hplus}-\ref{yminus}) neatly separating the effects of the processes with and without fermion number non-conservation, the goal which was attempted already in \cite{Shaposhnikov:2008pf} but not achieved at that time.

Now, we make these equations more realistic, accounting for the presence of charged fermions in the plasma, equilibrium character of electroweak reactions, and eventually sphaleron transitions. For this end we introduce leptonic numbers $\Delta L_\alpha$ of every generation, being a sum of asymmetries in neutrinos and charged leptons, integrated over momentum, and consider $\Delta_\alpha = \Delta L_\alpha -1/3\Delta B$, where $\Delta B$ is the baryon asymmetry. Due to  weak interactions the asymmetries in neutrinos are rapidly transferred to charge leptons and distributed among different momenta, whereas when sphalerons are operating there is a rapid transfer of lepton number to baryons. However, the rate of $\Delta_\alpha$ change is proportional to HNL Yukawa couplings and corresponds to a slow process. A well defined procedure accounting for equilibrium character of weak reactions and electric neutrality of the plasma \cite{Khlebnikov:1996vj,Laine:1999wv,Nardi:2006fx,Bodeker:2014hqa}  allows to write kinetic equations for $\Delta_\alpha$ instead of $\rho_{\nu_\alpha}$ and  $\rho_{\bar{\nu}_\alpha}$:
\begin{align}
i \frac{d\Delta_\alpha}{dt}
&= - i \left[ 2 \frac{\mu_\alpha}{T} \int \frac{d^{3}k}{(2 \pi)^{3}} \Gamma_{\nu_\alpha} f_{f} (1-f_{f}) \right] \, \nonumber \\
    &+ i \int \frac{d^{3}k}{(2 \pi)^{3}} \left[ \, \text{\text{Tr}}[\tilde{\Gamma}_{\nu_\alpha} \, \rho_{\bar{N}}]
    -  \, \text{\text{Tr}}[\tilde{\Gamma}_{\nu_\alpha}^\ast \, \rho_{N}] \right],
\label{KEDeltaL}\\
i \, \frac{d\rho_{N}}{dt} 
&= [H_N, \rho_N]
    - \frac{i}{2} \, \{ \Gamma_{N} , \rho_{N} \} \nonumber \\
    &- \frac{i}{2} \, \sum_\alpha \tilde{\Gamma}_{N}^\alpha \, \left[ 2 \frac{\mu_\alpha}{T} f_{f} (1-f_{f}) \right],
\label{KEN2}\\
i \, \frac{d\rho_{\bar{N}}}{dt} 
&= [H_N^\ast, \rho_{\bar{N}}]
    - \frac{i}{2} \, \{ \Gamma_{N}^\ast , \rho_{\bar{N}} \} \nonumber \\
    &+ \frac{i}{2} \, \sum_\alpha (\tilde{\Gamma}_{N}^\alpha)^\ast \, \left[ 2 \frac{\mu_\alpha}{T} f_{f} (1-f_{f}) \right],
\label{KENbar2}
\end{align}
where $\mu_{\alpha}$ are the chemical potentials for $\Delta_{\alpha}$ and $f_{f} = 1/(e^{k/T}+1)$ is the distribution function for massless fermion.
In the Higgs phase for $m_{b} \lesssim T \lesssim m_{W}$, where $m_{b}$ and $m_{W}$ are the masses of b-quark and W boson, the chemical potentials are given by
\begin{align}
\begin{split}
\frac{\mu_{\alpha}}{T} &= \frac{6}{T^{3}} \left[ \omega_{\alpha \beta} \Delta_{\beta} + \frac{10}{69} \Delta B_{0} \right], \\
\omega_{\alpha \beta} &= \frac{1}{207}
\begin{pmatrix}
79 & 10 & 10 \\
10 & 79 & 10 \\
10 & 10 & 79
\end{pmatrix},
\end{split}
\end{align}
where $\Delta B_{0}$ is the freeze-out baryon number and $\omega_{\alpha \beta}$ is the susceptibility matrix. 
Note that $\Delta_\alpha$ are the momentum-independent quantities (neutrinos and charged leptons of a given generation are well in thermal equilibrium and thus  their number densities are described well by the Fermi distribution with the chemical potentials for lepton numbers), but the density matrix for HNLs does depend on $k$. 

Yet another effect coming from charged leptons is the flavour dependence of the neutrino dispersion relations and neutrino damping rates. It can be neglected for temperatures exceeding the $\tau$-lepton mass. 

These completes the discussion of kinetic equations for HNLs and active flavours deeply in the Higgs phase, where the main contribution to active-sterile transition comes from the mixing terms contained in $\mathcal{L}_{int}$.

At the temperatures in the region of the electroweak crossover the structure of evolution equations remains the same, but a number of kinetic coefficients has to be modified. In particular, one has to add  the ``direct processes'' (in terminology of Ref. \cite{Ghiglieri:2016xye})  involving HNLs and active neutrinos. These processes occur with fermion number conservation and their contribution  does not vanish when  $\langle \Phi \rangle \to 0$. In fact, they dominate in baryogenesis around the sphaleron freeze-out in a part of the $\nu$MSM parameter space. The account for them results in the following modifications:
\begin{align}
 h_{+} &= {\cal K}(m_{h})\frac{T^{2}}{8 k} + \frac{2 \langle \Phi \rangle^2 E_\nu (E_N + k) (E_N + E_\nu)}{k E_N (4(E_N+E_\nu)^2 +\gamma_\nu^2)}, \\
 \gamma_{+} &= \gamma_{+}^{\text{direct}} + \frac{2 \langle \Phi \rangle^2 E_\nu (E_N + k) \gamma_\nu}{k E_N (4(E_N+E_\nu)^2 +\gamma_\nu^2)}, \label{eq:gPb2}
 \end{align}
where
 \begin{align}
 \gamma_+^{\text{direct}} &= {\cal K}(m_{h})  \frac{1}{E_N} \text{Im} \, \Pi_R + \gamma_{ph}, \label{eq:gPb_d}\\
{\cal K}(m_{h}) &= \frac{3}{\pi^{2}T^{3}} \int_{0}^{\infty}dp \, p^{2} f_{b}(E_{h}) (1+f_{b}(E_{h})), \label{eq:K_SF} \\
 \gamma_{ph} &= \frac{1}{E_{N}} \frac{m_{h}^{2} T}{32 \pi k} \ln \left\{ \frac{1+ e^{-\frac{m_{h}^{2}}{4 k T}}}{1-e^{-\frac{1}{T} (k+\frac{m_{h}^{2}}{4 k})}} \right\}, \label{eq:gPb_ph}
\end{align}
and $\text{Im}\,\Pi_R$ is the rate of the direct production of HNLs in the symmetric phase, which comes  mainly from $2 \leftrightarrow 2$ interactions. This contribution to $\gamma_+$ dies out in the Higgs phase; this is accounted for by a function  ${\cal K}(m_{h})$ \cite{Ghiglieri:2016xye}. The contribution to $\gamma_+$ from the Higgs decay to $N\nu$ is given by $\gamma_{ph}$ ~\cite{Ghiglieri:2016xye},   $f_{b}(\epsilon)=1/(e^{\epsilon/T}-1)$ is the bosonic distribution function.

In the symmetric phase these expressions simplify a lot:
 \begin{align}
 h_+ &= \frac{T^2}{8 k}, \\
 h_- &= 0, \\
 \gamma_+ &= \frac{1}{E_N} \text{Im} \, \Pi_R, \label{eq:gPs}\\
 \gamma_- &= 0,
\end{align}
where $h_+$ is nothing but the Weldon high-temperature correction \cite{Weldon:1982bn}. 

The last point is the modification of the susceptibility matrix in the symmetric phase, where sphalerons are in thermal equilibrium\footnote{The kinetic equations (\ref{eq:KEnu}-\ref{eq:KENbar}) should be modified and supplemented by an extra equation for baryon number when the rate of sphaleron transitions becomes smaller than the HNL equilibration rates \cite{Burnier:2005hp}.} :
\begin{align}
\begin{split}
\frac{\mu_{\alpha}}{T} &= \frac{6}{T^{3}} \, \omega_{\alpha \beta} \Delta_{\beta}, \\
\omega_{\alpha \beta} &= \frac{1}{711}
\begin{pmatrix}
257 & 20 & 20 \\
20 & 257 & 20 \\
20 & 20 & 257
\end{pmatrix}.
\end{split}
\end{align}
This formula is extracted from \cite{Khlebnikov:1996vj}. For determination of precise  temperature dependence of  $\omega_{\alpha \beta}$ over the electroweak crossover see Ref.~\cite{Ghiglieri:2016xye}. The relation between  $\Xi$  defined in Ref.~\cite{Ghiglieri:2016xye} and our $\omega_{\alpha \beta} $ reads $\omega_{\alpha \beta}   = \frac{1}{6 T^2} \Xi^{-1}$.

The set of equations derived in this section allows to follow the system from very high to sufficiently small temperatures $T\simeq 1$ GeV, and address both baryon asymmetry generation around $130$ GeV and late time lepton asymmetry production. At even smaller temperature one should take into account the flavour dependence of neutrino degrees of freedom in the medium, and include the decays and inverse decays of HNL which were omitted in our equations. 

\section{Thermal equilibrium and approximately conserved numbers}
\label{rates}
The kinetic equations (\ref{eq:KEnu}-\ref{eq:KENbar}) allow to address the question of existence of approximately conserved quantum numbers. Let us consider two combinations of the HNL and active flavour asymmetries, 
\begin{align}
L_{\pm}=  \Delta_{N} \mp \sum_{\alpha} \Delta_{\alpha}~,
\label{lpm}
\end{align}
where
\begin{align}
\Delta_{N} = \left[ \int \frac{d^{3}k}{(2 \pi)^{3}}\text{Tr}(\rho_{N} - \rho_{\bar{N}}) \right].
\end{align}

Making the corresponding linear combinations of Eqs.~(\ref{eq:KEnu}-\ref{eq:KENbar}), we find:
\begin{align}
 &\frac{d}{dt} L_- = - 2 \int \frac{d^{3}k}{(2 \pi)^{3}}  \gamma_{-} \sum_{\alpha} \biggl[ h_{\alpha 2} h_{\alpha 2}^{\ast} (\rho_{N, 11} - \rho_{\bar{N}, 11}) \nonumber \\
 & \hspace{50pt} + h_{\alpha 3} h_{\alpha 3}^{\ast} (\rho_{N, 22} - \rho_{\bar{N}, 22}) \nonumber \\
 & \hspace{50pt} - 2 \text{Re} (h_{\alpha 2} h_{\alpha 3}^{\ast}) (\text{Re} \rho_{N, 12} - \text{Re} \rho_{\bar{N}, 12}) \nonumber \\
 & \hspace{50pt} + 2 \text{Im} (h_{\alpha 2} h_{\alpha 3}^{\ast}) (\text{Im} \rho_{N, 12} + \text{Im} \rho_{\bar{N}, 12}) \nonumber \\
 & \hspace{50pt} + (h_{\alpha 2} h_{\alpha 2}^{\ast} + h_{\alpha 3} h_{\alpha 3}^{\ast}) \, \left[ 2 \frac{\mu_\alpha}{T} f_{f} (1-f_{f}) \right] \biggl],
\end{align}
\begin{align}
 &\frac{d}{dt} L_+ = - 2 \int \frac{d^{3}k}{(2 \pi)^{3}} \gamma_{+} \sum_{\alpha}  \biggl[ h_{\alpha 2} h_{\alpha 2}^{\ast} (\rho_{N, 11} - \rho_{\bar{N}, 11}) \nonumber \\
 & \hspace{50pt} + h_{\alpha 3} h_{\alpha 3}^{\ast} (\rho_{N, 22} - \rho_{\bar{N}, 22}) \nonumber \\
 & \hspace{50pt} - 2 \text{Re} (h_{\alpha 2} h_{\alpha 3}^{\ast}) (\text{Re} \rho_{N, 12} - \text{Re} \rho_{\bar{N}, 12}) \nonumber \\
 & \hspace{50pt} + 2 \text{Im} (h_{\alpha 2} h_{\alpha 3}^{\ast}) (\text{Im} \rho_{N, 12} + \text{Im} \rho_{\bar{N}, 12}) \nonumber \\
 & \hspace{50pt} - (h_{\alpha 2} h_{\alpha 2}^{\ast} + h_{\alpha 3} h_{\alpha 3}^{\ast}) \, \left[ 2 \frac{ \mu_\alpha}{T} f_{f} (1-f_{f}) \right]  \biggl].
\end{align}
The remarkable property of these relations is that the rates of the $L_\pm$ change is proportional to the corresponding $\gamma_\pm$ functions, which have very different behaviours as a function of temperature. In particular, the rate associated with $\gamma_+$ may never come into thermal equilibrium. 

\begin{figure}[!tb]
  \begin{center} 
    \includegraphics[clip, width=9cm]{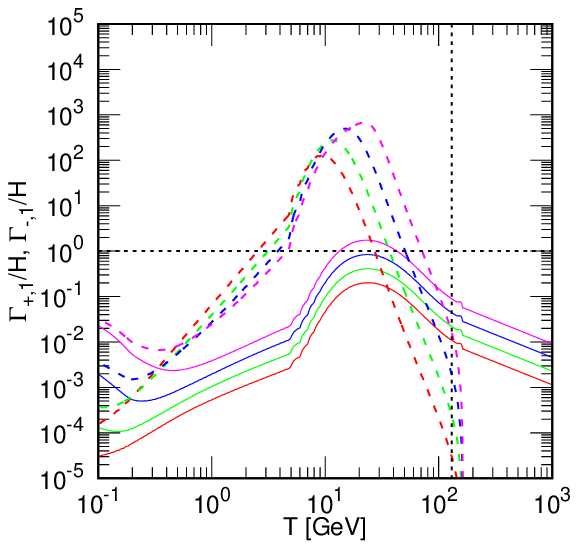} \\
    \includegraphics[clip, width=9cm]{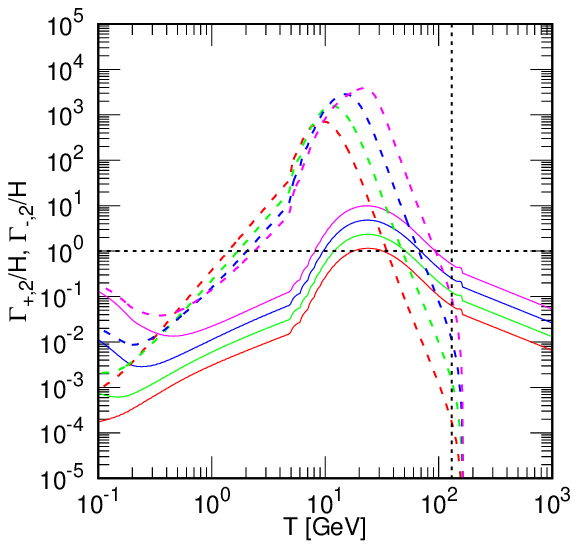} 
    \caption{Two eigen-values of momentum-averaged rates, $\Gamma_{+}$ (solid) and $\Gamma_{-}$ (dashed),  for $X_{\omega}=1$ in NH case. Red, green, blue and magenta lines correspond to $M=0.5, 1, 2$ and $4$~GeV, respectively. In this and all subsequent figures the vertical black dotted line shows the sphaleron freeze-out temperature 130 GeV, and on the horizontal black dotted line $\Gamma = H$.}
    \label{fig:GammaM_N}
  \end{center}
\end{figure}

\begin{figure}[!tb]
  \begin{center} 
    \includegraphics[clip, width=9cm]{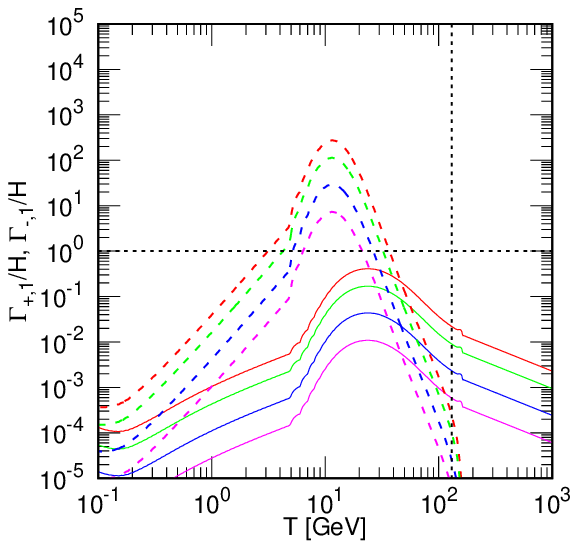} \\
    \includegraphics[clip, width=9cm]{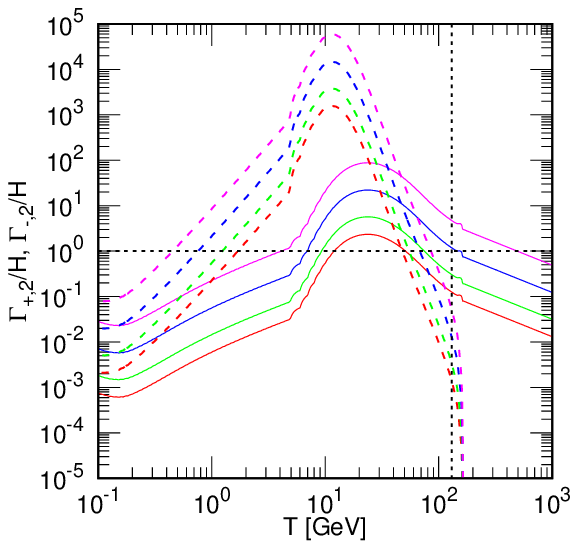} 
    \caption{Two eigen-values of momentum-averaged rates, $\Gamma_{+}$ (solid) and $\Gamma_{-}$ (dashed), for $M=1$~GeV in NH case. Red, green, blue and magenta lines correspond to $X_{\omega}=1, 2, 4$ and $8$, respectively. }
    \label{fig:GammaX_N}
  \end{center}
\end{figure}

\begin{figure}[!tb]
  \begin{center} 
    \includegraphics[clip, width=9cm]{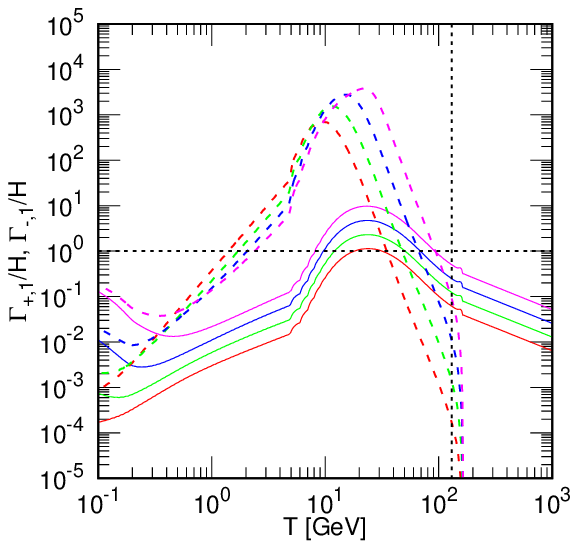} \\
    \includegraphics[clip, width=9cm]{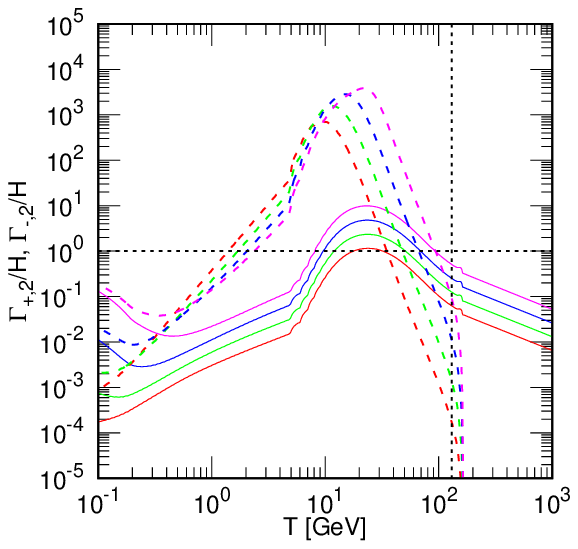} 
    \caption{Two eigen-values of momentum-averaged rates, $\Gamma_{+}$ (solid) and $\Gamma_{-}$ (dashed), for $X_{\omega}=1$ in IH case. Red, green, blue and magenta lines correspond to $M=0.5, 1, 2$ and $4$~GeV, respectively. }
    \label{fig:GammaM_I}
  \end{center}
\end{figure}

\begin{figure}[!tb]
  \begin{center} 
    \includegraphics[clip, width=9cm]{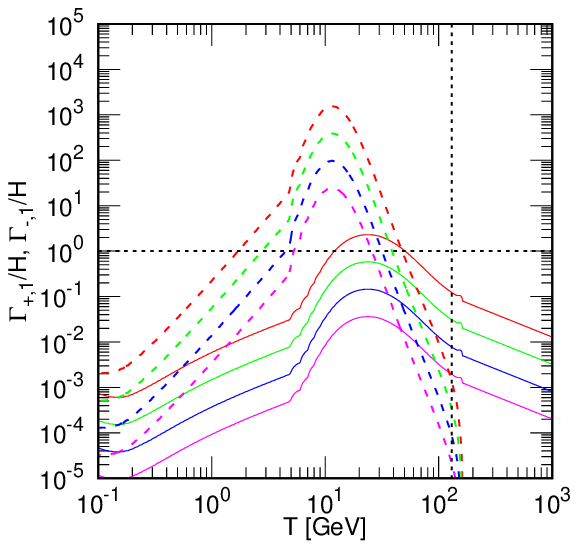} \\
    \includegraphics[clip, width=9cm]{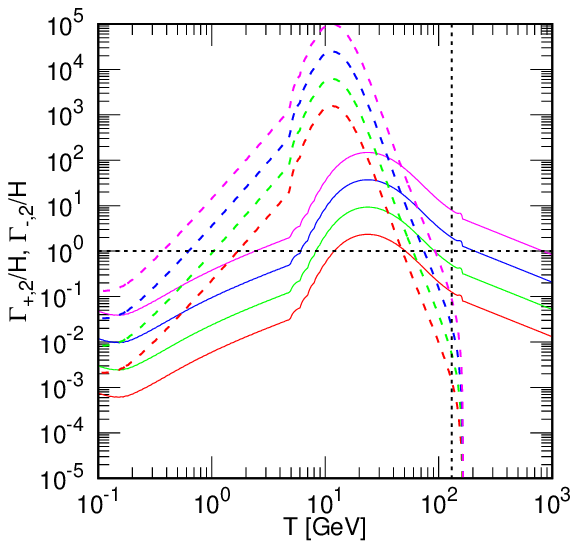} 
    \caption{Two eigen-values of momentum-averaged rates, $\Gamma_{+}$ (solid) and $\Gamma_{-}$ (dashed), for $M=1$~GeV in IH case. Red, green, blue and magenta lines correspond to $X_{\omega}=1, 2, 4$ and $8$, respectively.}
    \label{fig:GammaX_I}
  \end{center}
\end{figure}

We plot the ratio of the different HNL production rates to the Hubble rate $H$ in Figures~\ref{fig:GammaM_N} and \ref{fig:GammaX_N} for the normal hierarchy (NH) case and in Figures~\ref{fig:GammaM_I} and \ref{fig:GammaX_I} for the inverted hierarchy (IH). The $\Gamma_\pm$ rates defined in Eqs.~(\ref{ga+},\ref{ga-}) are $2\times2$ matrices, so we diagonalise each of them at any given temperature and denote by $\Gamma_{\pm,i}$ with $i=1,2$ marking the corresponding eigen-values. In addition, we average the rates over momentum with the use of the Fermi distribution function, $f_F$, 
\begin{align}
 \langle \Gamma\rangle=\frac{1}{n_F}\int \frac{d^{3}k}{(2 \pi)^{3}}f_F(k) \Gamma(k)~,
\end{align}
where $n_F$ is the fermion number density.

The rates depend on quite a number of parameters, the most important being the HNL mass $M$ and the imaginary part of a  complex mixing angle $\omega$ appearing in the Casas-Ibarra parametrisation of the HNL-neutrino mixing matrix \cite{Casas:2001sr}. The quantity $X_{\omega} \equiv \exp(\text{Im}\,\omega)$ shown in the figures is a free parameter which does not change the active neutrino masses (we fix them with the available neutrino data). The amplitude of Yukawa couplings scales as $F_{\alpha I} \propto X_{\omega}$ for large $\text{Im} \omega > 0$ or $F_{\alpha I} \propto X_{\omega}^{-1}$ for $\text{Im} \omega < 0$.  The parameter $X_{\omega} $  is related to the introduced previously $\epsilon$ in \cite{Shaposhnikov:2006nn,Shaposhnikov:2008pf} as $\epsilon = 1/X_\omega^2$. More specifically, the relation between Yukawa couplings and neutrino parameters, when $\Delta M$ is negligibly small,  is given by
\begin{align}
 \sum_{\alpha} h_{\alpha 2} h_{\alpha 2}^{\ast} &= \frac{(m_3+m_2)M}{2 v^{2}} X_\omega^{-2}, \label{eq:h22}\\
 \sum_{\alpha} h_{\alpha 3} h_{\alpha 3}^{\ast} &= \frac{(m_3+m_2)M}{2 v^{2}} X_\omega^2, \label{eq:h33}\\
 \sum_{\alpha} \text{Re} \, (h_{\alpha 2} h_{\alpha 3}^{\ast}) &= \frac{(m_3-m_2)M}{2 v^{2}}\cos(2 \text{Re}\,\omega), \label{eq:Rh23}\\
\sum_{\alpha} \text{Im} \, (h_{\alpha 2} h_{\alpha 3}^{\ast}) &= \frac{(m_3-m_2)M}{2 v^{2}}\sin(2\text{Re}\,\omega), \label{eq:Ih23}
\end{align}
where $m_{3}$ and $m_{2}$ are the heaviest and second-heaviest active neutrino masses and $v=174.1$~GeV is the Higgs expectation value at zero temperature.
It is the largest eigenvalue of  $\Gamma_\pm$  which determines the approach to thermal equilibrium for quantum numbers $L_\pm$. In general, the equilibration rates are larger for larger HNL masses, and smaller if $X_{\omega}$ is close to one.

In Figures~\ref{fig:GammaM_N} and~\ref{fig:GammaM_I}  we show the dependence of the rates on temperature for different HNL masses. 

In Figures \ref{fig:GammaX_N}  and  \ref{fig:GammaX_I}  the behaviour of the rates as a function of temperature for different values of  $X_{\omega}$ is shown. The large (or small)  $X_{\omega}$ increases the magnitude of Yukawa couplings, leading to faster equilibration. 

Let us comment on the splitting between the eigen-values of the rates. From Eqs.~(\ref{eq:h22}-\ref{eq:Ih23}) for large (or small) $X_{\omega}$ the hierarchy between the diagonal components of the rates (\ref{ga+},\ref{ga-},\ref{yplus},\ref{yminus}) gets large, which is reflected in the splitting of eigen-values. Note that the off-diagonal  components  of them  are independent on $X_\omega$. The splitting is clearly visible in the NH case even if $X_{\omega} \simeq 1$ due to the presence of non-diagonal terms (\ref{eq:Rh23},\ref{eq:Ih23}) which are of the order of  atmospheric neutrino mass scale $m_{\text{atm}} \simeq 5 \times 10^{-2} ~\text{eV}$. For the IH case the eigen-values of the rates  at $X_{\omega} \simeq 1$ are almost degenerate  due to the smallness of the off-diagonal components suppressed by $(m_{\text{sol}}/m_{\text{atm}})^{2}$ where the solar neutrino mass scale is $m_{\text{sol}} \simeq 9 \times 10^{-3}~ \text{eV}$.
  
In the symmetric phase at $T>160$ GeV, $\Gamma_{-}=0$ and the lepton number $L_{-}$ is conserved.  The generated asymmetry in the active neutrino sector is the same as that in the HNL sector with an opposite sign. In the temperature range relevant for baryogenesis, i.e. above the sphaleron freeze-out $T\simeq 130$ GeV but below $T=160$ GeV  both  $\Gamma_{-}$ and $\Gamma_{+}$ rates are present. However, the inspection of the figures shows that  the  rate $\Gamma_{+}$ dominates over $\Gamma_{-}$, meaning that the discussion above is approximately valid, with small corrections from ``-'' contributions. As a result the previous computations of the baryon asymmetry performed in  \cite{Asaka:2005pn,Shaposhnikov:2006nn,Shaposhnikov:2008pf,Canetti:2010aw,Asaka:2010kk,Asaka:2011wq,Canetti:2012vf,Canetti:2012kh,Shuve:2014zua,Abada:2015rta,Hernandez:2016kel,Drewes:2016gmt}  neglecting $\Gamma_{-}$ are legitimate. 

When the temperature goes down below $T\simeq 130$ GeV the generation of the baryon asymmetry stops but of the lepton asymmetry continues \cite{Shaposhnikov:2008pf}. Eventually, the rate  $\Gamma_{-}$  starts  to dominate over  $\Gamma_{+}$.  The HNLs enter in thermal equilibrium at temperature $T_\text{in}$ corresponding to the intersection of the largest rate with the horizontal line $\Gamma/H=1$, and go out of thermal equilibrium at $T=T_\text{out}<T_\text{in}$ found in a similar way. Typically, for the set of masses considered, the highest rate is achieved at $T\simeq 10-20$ GeV. The maximum of the rate $\Gamma_-$ always exceeds the rate of the Universe expansion, as has been already demonstrated in  \cite{Shaposhnikov:2008pf,Canetti:2012kh,Ghisoiu:2014ena,Ghiglieri:2016xye}.

The most interesting range of parameters that can potentially lead to generation of large leptonic asymmetries at $T\simeq T_\text{in}$ corresponds to relatively small HNL masses and $X_\omega \simeq 1$. Indeed, the Figures~ \ref{fig:GammaM_N}-\ref{fig:GammaX_I}  show that the maximum of the ratio $\Gamma_{+,2}/H$ does not exceed ${\cal O}(1)$ for $M \lesssim 2$ GeV and $X_\omega \simeq 1$. At the same time, this ratio is close to $1$ for these parameters, meaning that the large asymmetry in $L_+$ can potentially be generated but protected from washout until small temperature.   
The character of equilibrium which is led by the "-" reactions at smaller temperatures but has an (effective) $L_{+}$ conservation will ensure the relation between the asymmetries in active flavours and HNLs given by\footnote{This is calculated following the similar analysis in  ~\cite{Khlebnikov:1996vj}.}
\begin{align}
\frac{\Delta_{N}}{\sum_{\alpha} \Delta_{\alpha}} \approx - \frac{22}{69} \, .
\end{align}
Note that only the asymmetries in active flavours contribute to the resonant DM production. 
We leave the analysis of produced lepton asymmetries for a future publication. 


\section{Conclusions}
\label{concl}
In this paper we have derived kinetic equations for low scale leptogenesis, accounting for helicity effects in HNL interactions with active flavours. They are valid both for high temperatures around the sphaleron freeze-out (to describe the baryogenesis), and for smaller temperatures to account for lepton number generation.

We showed that the total lepton number can be considered as approximately conserved in the certain domain of $\nu$MSM parameters, and thus can potentially be generated at  $T \simeq T_\text{in}$ and survive until small temperatures making the resonant production of DM possible.  It is still remains to be seen whether sufficiently large lepton asymmetry is indeed generated, as these depends on CP-violating effects. The work in this direction is in progress.

Quite remarkably, the requirement of existence of approximately conserved leptonic number makes the region of small HNL masses preferable,  allowing to search for HNLs at SHiP  or SHiP --  like experiments.  At the same time, it  lies close to the lower bound coming from the requirement to explain neutrino masses via see-saw, urging to design the experiment in this mass region with highest possible sensitivity.


\section*{Acknowledgements}

This work was supported by the ERC-AdG-2015 grant 694896. We are grateful to Mikko Laine for many illuminating discussions. We thank A. Boyarsky, O. Ruchayskiy and I. Timiryasov for helpful comments. The work of MS was supported partially by the Swiss National Science Foundation. 

Once the current manuscript had been finalised, we thank M.~Laine for sharing with us a draft  by himself and J. Ghiglieri, in which helicity-flipping and conserving rates similar to ours are discussed at $T > 130$~GeV. Their remarks on our manuscript are  appreciated.



\begin{thebibliography}{10}
\expandafter\ifx\csname url\endcsname\relax
  \def\url#1{\texttt{#1}}\fi
\expandafter\ifx\csname urlprefix\endcsname\relax\def\urlprefix{URL }\fi
\expandafter\ifx\csname href\endcsname\relax
  \def\href#1#2{#2} \def\path#1{#1}\fi

\bibitem{Bezrukov:2012sa}
F.~Bezrukov, M.~{\relax Yu}. Kalmykov, B.~A. Kniehl, M.~Shaposhnikov, {Higgs
  Boson Mass and New Physics}, JHEP 10 (2012) 140.
\newblock \href {http://arxiv.org/abs/1205.2893} {\path{arXiv:1205.2893}},
  \href {http://dx.doi.org/10.1007/JHEP10(2012)140}
  {\path{doi:10.1007/JHEP10(2012)140}}.

\bibitem{Buttazzo:2013uya}
D.~Buttazzo, G.~Degrassi, P.~P. Giardino, G.~F. Giudice, F.~Sala, A.~Salvio,
  A.~Strumia, {Investigating the near-criticality of the Higgs boson}, JHEP 12
  (2013) 089.
\newblock \href {http://arxiv.org/abs/1307.3536} {\path{arXiv:1307.3536}},
  \href {http://dx.doi.org/10.1007/JHEP12(2013)089}
  {\path{doi:10.1007/JHEP12(2013)089}}.

\bibitem{Bednyakov:2015sca}
A.~V. Bednyakov, B.~A. Kniehl, A.~F. Pikelner, O.~L. Veretin, {Stability of the
  Electroweak Vacuum: Gauge Independence and Advanced Precision}, Phys. Rev.
  Lett. 115~(20) (2015) 201802.
\newblock \href {http://arxiv.org/abs/1507.08833} {\path{arXiv:1507.08833}},
  \href {http://dx.doi.org/10.1103/PhysRevLett.115.201802}
  {\path{doi:10.1103/PhysRevLett.115.201802}}.

\bibitem{Asaka:2005an}
T.~Asaka, S.~Blanchet, M.~Shaposhnikov, {The nuMSM, dark matter and neutrino
  masses}, Phys. Lett. B631 (2005) 151--156.
\newblock \href {http://arxiv.org/abs/hep-ph/0503065}
  {\path{arXiv:hep-ph/0503065}}, \href
  {http://dx.doi.org/10.1016/j.physletb.2005.09.070}
  {\path{doi:10.1016/j.physletb.2005.09.070}}.

\bibitem{Asaka:2005pn}
T.~Asaka, M.~Shaposhnikov, {The nuMSM, dark matter and baryon asymmetry of the
  universe}, Phys. Lett. B620 (2005) 17--26.
\newblock \href {http://arxiv.org/abs/hep-ph/0505013}
  {\path{arXiv:hep-ph/0505013}}, \href
  {http://dx.doi.org/10.1016/j.physletb.2005.06.020}
  {\path{doi:10.1016/j.physletb.2005.06.020}}.

\bibitem{Agashe:2014kda}
K.~A. Olive, et~al., {Review of Particle Physics}, Chin. Phys. C38 (2014)
  090001.
\newblock \href {http://dx.doi.org/10.1088/1674-1137/38/9/090001}
  {\path{doi:10.1088/1674-1137/38/9/090001}}.

\bibitem{Dodelson:1993je}
S.~Dodelson, L.~M. Widrow, {Sterile-neutrinos as dark matter}, Phys. Rev. Lett.
  72 (1994) 17--20.
\newblock \href {http://arxiv.org/abs/hep-ph/9303287}
  {\path{arXiv:hep-ph/9303287}}, \href
  {http://dx.doi.org/10.1103/PhysRevLett.72.17}
  {\path{doi:10.1103/PhysRevLett.72.17}}.

\bibitem{Shi:1998km}
X.-D. Shi, G.~M. Fuller, {A New dark matter candidate: Nonthermal sterile
  neutrinos}, Phys. Rev. Lett. 82 (1999) 2832--2835.
\newblock \href {http://arxiv.org/abs/astro-ph/9810076}
  {\path{arXiv:astro-ph/9810076}}, \href
  {http://dx.doi.org/10.1103/PhysRevLett.82.2832}
  {\path{doi:10.1103/PhysRevLett.82.2832}}.

\bibitem{Akhmedov:1998qx}
E.~K. Akhmedov, V.~A. Rubakov, A.~{\relax Yu}. Smirnov, {Baryogenesis via
  neutrino oscillations}, Phys. Rev. Lett. 81 (1998) 1359--1362.
\newblock \href {http://arxiv.org/abs/hep-ph/9803255}
  {\path{arXiv:hep-ph/9803255}}, \href
  {http://dx.doi.org/10.1103/PhysRevLett.81.1359}
  {\path{doi:10.1103/PhysRevLett.81.1359}}.

\bibitem{Shaposhnikov:2006nn}
M.~Shaposhnikov, {A Possible symmetry of the nuMSM}, Nucl. Phys. B763 (2007)
  49--59.
\newblock \href {http://arxiv.org/abs/hep-ph/0605047}
  {\path{arXiv:hep-ph/0605047}}, \href
  {http://dx.doi.org/10.1016/j.nuclphysb.2006.11.003}
  {\path{doi:10.1016/j.nuclphysb.2006.11.003}}.

\bibitem{Shaposhnikov:2008pf}
M.~Shaposhnikov, {The nuMSM, leptonic asymmetries, and properties of singlet
  fermions}, JHEP 08 (2008) 008.
\newblock \href {http://arxiv.org/abs/0804.4542} {\path{arXiv:0804.4542}},
  \href {http://dx.doi.org/10.1088/1126-6708/2008/08/008}
  {\path{doi:10.1088/1126-6708/2008/08/008}}.

\bibitem{Canetti:2010aw}
L.~Canetti, M.~Shaposhnikov, {Baryon Asymmetry of the Universe in the NuMSM},
  JCAP 1009 (2010) 001.
\newblock \href {http://arxiv.org/abs/1006.0133} {\path{arXiv:1006.0133}},
  \href {http://dx.doi.org/10.1088/1475-7516/2010/09/001}
  {\path{doi:10.1088/1475-7516/2010/09/001}}.

\bibitem{Asaka:2010kk}
T.~Asaka, H.~Ishida, {Flavour Mixing of Neutrinos and Baryon Asymmetry of the
  Universe}, Phys. Lett. B692 (2010) 105--113.
\newblock \href {http://arxiv.org/abs/1004.5491} {\path{arXiv:1004.5491}},
  \href {http://dx.doi.org/10.1016/j.physletb.2010.07.016}
  {\path{doi:10.1016/j.physletb.2010.07.016}}.

\bibitem{Asaka:2011wq}
T.~Asaka, S.~Eijima, H.~Ishida, {Kinetic Equations for Baryogenesis via Sterile
  Neutrino Oscillation}, JCAP 1202 (2012) 021.
\newblock \href {http://arxiv.org/abs/1112.5565} {\path{arXiv:1112.5565}},
  \href {http://dx.doi.org/10.1088/1475-7516/2012/02/021}
  {\path{doi:10.1088/1475-7516/2012/02/021}}.

\bibitem{Canetti:2012vf}
L.~Canetti, M.~Drewes, M.~Shaposhnikov, {Sterile Neutrinos as the Origin of
  Dark and Baryonic Matter}, Phys. Rev. Lett. 110~(6) (2013) 061801.
\newblock \href {http://arxiv.org/abs/1204.3902} {\path{arXiv:1204.3902}},
  \href {http://dx.doi.org/10.1103/PhysRevLett.110.061801}
  {\path{doi:10.1103/PhysRevLett.110.061801}}.

\bibitem{Canetti:2012kh}
L.~Canetti, M.~Drewes, T.~Frossard, M.~Shaposhnikov, {Dark Matter, Baryogenesis
  and Neutrino Oscillations from Right Handed Neutrinos}, Phys. Rev. D87 (2013)
  093006.
\newblock \href {http://arxiv.org/abs/1208.4607} {\path{arXiv:1208.4607}},
  \href {http://dx.doi.org/10.1103/PhysRevD.87.093006}
  {\path{doi:10.1103/PhysRevD.87.093006}}.

\bibitem{Shuve:2014zua}
B.~Shuve, I.~Yavin, {Baryogenesis through Neutrino Oscillations: A Unified
  Perspective}, Phys. Rev. D89~(7) (2014) 075014.
\newblock \href {http://arxiv.org/abs/1401.2459} {\path{arXiv:1401.2459}},
  \href {http://dx.doi.org/10.1103/PhysRevD.89.075014}
  {\path{doi:10.1103/PhysRevD.89.075014}}.

\bibitem{Abada:2015rta}
A.~Abada, G.~Arcadi, V.~Domcke, M.~Lucente, {Lepton number violation as a key
  to low-scale leptogenesis}, JCAP 1511~(11) (2015) 041.
\newblock \href {http://arxiv.org/abs/1507.06215} {\path{arXiv:1507.06215}},
  \href {http://dx.doi.org/10.1088/1475-7516/2015/11/041}
  {\path{doi:10.1088/1475-7516/2015/11/041}}.

\bibitem{Hernandez:2016kel}
P.~Hernández, M.~Kekic, J.~López-Pavón, J.~Racker, J.~Salvado, {Testable
  Baryogenesis in Seesaw Models}, JHEP 08 (2016) 157.
\newblock \href {http://arxiv.org/abs/1606.06719} {\path{arXiv:1606.06719}},
  \href {http://dx.doi.org/10.1007/JHEP08(2016)157}
  {\path{doi:10.1007/JHEP08(2016)157}}.

\bibitem{Drewes:2016gmt}
M.~Drewes, B.~Garbrecht, D.~Gueter, J.~Klaric, {Leptogenesis from Oscillations
  of Heavy Neutrinos with Large Mixing Angles}, JHEP 12 (2016) 150.
\newblock \href {http://arxiv.org/abs/1606.06690} {\path{arXiv:1606.06690}},
  \href {http://dx.doi.org/10.1007/JHEP12(2016)150}
  {\path{doi:10.1007/JHEP12(2016)150}}.

\bibitem{Bezrukov:2007ep}
F.~L. Bezrukov, M.~Shaposhnikov, {The Standard Model Higgs boson as the
  inflaton}, Phys. Lett. B659 (2008) 703--706.
\newblock \href {http://arxiv.org/abs/0710.3755} {\path{arXiv:0710.3755}},
  \href {http://dx.doi.org/10.1016/j.physletb.2007.11.072}
  {\path{doi:10.1016/j.physletb.2007.11.072}}.

\bibitem{Bezrukov:2008ut}
F.~Bezrukov, D.~Gorbunov, M.~Shaposhnikov, {On initial conditions for the Hot
  Big Bang}, JCAP 0906 (2009) 029.
\newblock \href {http://arxiv.org/abs/0812.3622} {\path{arXiv:0812.3622}},
  \href {http://dx.doi.org/10.1088/1475-7516/2009/06/029}
  {\path{doi:10.1088/1475-7516/2009/06/029}}.

\bibitem{GarciaBellido:2008ab}
J.~Garcia-Bellido, D.~G. Figueroa, J.~Rubio, {Preheating in the Standard Model
  with the Higgs-Inflaton coupled to gravity}, Phys. Rev. D79 (2009) 063531.
\newblock \href {http://arxiv.org/abs/0812.4624} {\path{arXiv:0812.4624}},
  \href {http://dx.doi.org/10.1103/PhysRevD.79.063531}
  {\path{doi:10.1103/PhysRevD.79.063531}}.

\bibitem{Bezrukov:2014ipa}
F.~Bezrukov, J.~Rubio, M.~Shaposhnikov, {Living beyond the edge: Higgs
  inflation and vacuum metastability}, Phys. Rev. D92~(8) (2015) 083512.
\newblock \href {http://arxiv.org/abs/1412.3811} {\path{arXiv:1412.3811}},
  \href {http://dx.doi.org/10.1103/PhysRevD.92.083512}
  {\path{doi:10.1103/PhysRevD.92.083512}}.

\bibitem{Kuzmin:1985mm}
V.~A. Kuzmin, V.~A. Rubakov, M.~E. Shaposhnikov, {On the Anomalous Electroweak
  Baryon Number Nonconservation in the Early Universe}, Phys. Lett. B155 (1985)
  36.
\newblock \href {http://dx.doi.org/10.1016/0370-2693(85)91028-7}
  {\path{doi:10.1016/0370-2693(85)91028-7}}.

\bibitem{Abazajian:2001nj}
K.~Abazajian, G.~M. Fuller, M.~Patel, {Sterile neutrino hot, warm, and cold
  dark matter}, Phys. Rev. D64 (2001) 023501.
\newblock \href {http://arxiv.org/abs/astro-ph/0101524}
  {\path{arXiv:astro-ph/0101524}}, \href
  {http://dx.doi.org/10.1103/PhysRevD.64.023501}
  {\path{doi:10.1103/PhysRevD.64.023501}}.

\bibitem{Asaka:2006rw}
T.~Asaka, M.~Laine, M.~Shaposhnikov, {On the hadronic contribution to sterile
  neutrino production}, JHEP 06 (2006) 053.
\newblock \href {http://arxiv.org/abs/hep-ph/0605209}
  {\path{arXiv:hep-ph/0605209}}, \href
  {http://dx.doi.org/10.1088/1126-6708/2006/06/053}
  {\path{doi:10.1088/1126-6708/2006/06/053}}.

\bibitem{Asaka:2006nq}
T.~Asaka, M.~Laine, M.~Shaposhnikov, {Lightest sterile neutrino abundance
  within the nuMSM}, JHEP 01 (2007) 091, [Erratum: JHEP02,028(2015)].
\newblock \href {http://arxiv.org/abs/hep-ph/0612182}
  {\path{arXiv:hep-ph/0612182}}, \href
  {http://dx.doi.org/10.1088/1126-6708/2007/01/091, 10.1007/JHEP02(2015)028}
  {\path{doi:10.1088/1126-6708/2007/01/091, 10.1007/JHEP02(2015)028}}.

\bibitem{Laine:2008pg}
M.~Laine, M.~Shaposhnikov, {Sterile neutrino dark matter as a consequence of
  nuMSM-induced lepton asymmetry}, JCAP 0806 (2008) 031.
\newblock \href {http://arxiv.org/abs/0804.4543} {\path{arXiv:0804.4543}},
  \href {http://dx.doi.org/10.1088/1475-7516/2008/06/031}
  {\path{doi:10.1088/1475-7516/2008/06/031}}.

\bibitem{Ghiglieri:2015jua}
J.~Ghiglieri, M.~Laine, {Improved determination of sterile neutrino dark matter
  spectrum}, JHEP 11 (2015) 171.
\newblock \href {http://arxiv.org/abs/1506.06752} {\path{arXiv:1506.06752}},
  \href {http://dx.doi.org/10.1007/JHEP11(2015)171}
  {\path{doi:10.1007/JHEP11(2015)171}}.

\bibitem{Ghisoiu:2014ena}
I.~Ghisoiu, M.~Laine, {Right-handed neutrino production rate at $T > 160$ GeV},
  JCAP 1412~(12) (2014) 032.
\newblock \href {http://arxiv.org/abs/1411.1765} {\path{arXiv:1411.1765}},
  \href {http://dx.doi.org/10.1088/1475-7516/2014/12/032}
  {\path{doi:10.1088/1475-7516/2014/12/032}}.

\bibitem{Ghiglieri:2016xye}
J.~Ghiglieri, M.~Laine, {Neutrino dynamics below the electroweak crossover},
  JCAP 1607~(07) (2016) 015.
\newblock \href {http://arxiv.org/abs/1605.07720} {\path{arXiv:1605.07720}},
  \href {http://dx.doi.org/10.1088/1475-7516/2016/07/015}
  {\path{doi:10.1088/1475-7516/2016/07/015}}.

\bibitem{Boyarsky:2008ju}
A.~Boyarsky, O.~Ruchayskiy, D.~Iakubovskyi, {A Lower bound on the mass of Dark
  Matter particles}, JCAP 0903 (2009) 005.
\newblock \href {http://arxiv.org/abs/0808.3902} {\path{arXiv:0808.3902}},
  \href {http://dx.doi.org/10.1088/1475-7516/2009/03/005}
  {\path{doi:10.1088/1475-7516/2009/03/005}}.

\bibitem{Gorbunov:2008ka}
D.~Gorbunov, A.~Khmelnitsky, V.~Rubakov, {Constraining sterile neutrino dark
  matter by phase-space density observations}, JCAP 0810 (2008) 041.
\newblock \href {http://arxiv.org/abs/0808.3910} {\path{arXiv:0808.3910}},
  \href {http://dx.doi.org/10.1088/1475-7516/2008/10/041}
  {\path{doi:10.1088/1475-7516/2008/10/041}}.

\bibitem{Roy:2010xq}
A.~Roy, M.~Shaposhnikov, {Resonant production of the sterile neutrino dark
  matter and fine-tunings in the $\nu$MSM}, Phys. Rev. D82 (2010) 056014.
\newblock \href {http://arxiv.org/abs/1006.4008} {\path{arXiv:1006.4008}},
  \href {http://dx.doi.org/10.1103/PhysRevD.82.056014}
  {\path{doi:10.1103/PhysRevD.82.056014}}.

\bibitem{Blondel:2014bra}
A.~Blondel, E.~Graverini, N.~Serra, M.~Shaposhnikov,
  \href{https://inspirehep.net/record/1328783/files/arXiv:1411.5230.pdf}{{Search
  for Heavy Right Handed Neutrinos at the FCC-ee}}, in: {37th International
  Conference on High Energy Physics (ICHEP 2014) Valencia, Spain, July 2-9,
  2014}, 2014.
\newblock \href {http://arxiv.org/abs/1411.5230} {\path{arXiv:1411.5230}}.

\bibitem{Shaposhnikov:2006xi}
M.~Shaposhnikov, I.~Tkachev, {The nuMSM, inflation, and dark matter}, Phys.
  Lett. B639 (2006) 414--417.
\newblock \href {http://arxiv.org/abs/hep-ph/0604236}
  {\path{arXiv:hep-ph/0604236}}, \href
  {http://dx.doi.org/10.1016/j.physletb.2006.06.063}
  {\path{doi:10.1016/j.physletb.2006.06.063}}.

\bibitem{Kusenko:2006rh}
A.~Kusenko, {Sterile neutrinos, dark matter, and the pulsar velocities in
  models with a Higgs singlet}, Phys. Rev. Lett. 97 (2006) 241301.
\newblock \href {http://arxiv.org/abs/hep-ph/0609081}
  {\path{arXiv:hep-ph/0609081}}, \href
  {http://dx.doi.org/10.1103/PhysRevLett.97.241301}
  {\path{doi:10.1103/PhysRevLett.97.241301}}.

\bibitem{Bezrukov:2011sz}
F.~Bezrukov, D.~Gorbunov, M.~Shaposhnikov, {Late and early time phenomenology
  of Higgs-dependent cutoff}, JCAP 1110 (2011) 001.
\newblock \href {http://arxiv.org/abs/1106.5019} {\path{arXiv:1106.5019}},
  \href {http://dx.doi.org/10.1088/1475-7516/2011/10/001}
  {\path{doi:10.1088/1475-7516/2011/10/001}}.

\bibitem{Hambye:2016sby}
T.~Hambye, D.~Teresi, {Higgs doublet decay as the origin of the baryon
  asymmetry}, Phys. Rev. Lett. 117~(9) (2016) 091801.
\newblock \href {http://arxiv.org/abs/1606.00017} {\path{arXiv:1606.00017}},
  \href {http://dx.doi.org/10.1103/PhysRevLett.117.091801}
  {\path{doi:10.1103/PhysRevLett.117.091801}}.

\bibitem{Khlebnikov:1996vj}
S.~{\relax Yu}. Khlebnikov, M.~E. Shaposhnikov, {Melting of the Higgs vacuum:
  Conserved numbers at high temperature}, Phys. Lett. B387 (1996) 817--822.
\newblock \href {http://arxiv.org/abs/hep-ph/9607386}
  {\path{arXiv:hep-ph/9607386}}, \href
  {http://dx.doi.org/10.1016/0370-2693(96)01116-1}
  {\path{doi:10.1016/0370-2693(96)01116-1}}.

\bibitem{Laine:1999wv}
M.~Laine, M.~E. Shaposhnikov, {A Remark on sphaleron erasure of baryon
  asymmetry}, Phys. Rev. D61 (2000) 117302.
\newblock \href {http://arxiv.org/abs/hep-ph/9911473}
  {\path{arXiv:hep-ph/9911473}}, \href
  {http://dx.doi.org/10.1103/PhysRevD.61.117302}
  {\path{doi:10.1103/PhysRevD.61.117302}}.

\bibitem{Nardi:2006fx}
E.~Nardi, Y.~Nir, E.~Roulet, J.~Racker, {The Importance of flavor in
  leptogenesis}, JHEP 01 (2006) 164.
\newblock \href {http://arxiv.org/abs/hep-ph/0601084}
  {\path{arXiv:hep-ph/0601084}}, \href
  {http://dx.doi.org/10.1088/1126-6708/2006/01/164}
  {\path{doi:10.1088/1126-6708/2006/01/164}}.

\bibitem{Bodeker:2014hqa}
D.~Bodeker, M.~Laine, {Kubo relations and radiative corrections for lepton
  number washout}, JCAP 1405 (2014) 041.
\newblock \href {http://arxiv.org/abs/1403.2755} {\path{arXiv:1403.2755}},
  \href {http://dx.doi.org/10.1088/1475-7516/2014/05/041}
  {\path{doi:10.1088/1475-7516/2014/05/041}}.

\bibitem{Sigl:1992fn}
G.~Sigl, G.~Raffelt, {General kinetic description of relativistic mixed
  neutrinos}, Nucl. Phys. B406 (1993) 423--451.
\newblock \href {http://dx.doi.org/10.1016/0550-3213(93)90175-O}
  {\path{doi:10.1016/0550-3213(93)90175-O}}.

\bibitem{Wright:2013sv}
J.~G. Wright, B.~S. Shastry, {DiracQ: A Quantum Many-Body Physics
  Package.~}\href {http://arxiv.org/abs/1301.4494} {\path{arXiv:1301.4494}}.

\bibitem{Notzold:1987ik}
D.~Notzold, G.~Raffelt, {Neutrino Dispersion at Finite Temperature and
  Density}, Nucl. Phys. B307 (1988) 924.
\newblock \href {http://dx.doi.org/10.1016/0550-3213(88)90113-7}
  {\path{doi:10.1016/0550-3213(88)90113-7}}.

\bibitem{Morales:1999ia}
J.~Morales, C.~Quimbay, F.~Fonseca, {Fermionic dispersion relations at finite
  temperature and nonvanishing chemical potentials in the minimal standard
  model}, Nucl. Phys. B560 (1999) 601--616.
\newblock \href {http://arxiv.org/abs/hep-ph/9906207}
  {\path{arXiv:hep-ph/9906207}}, \href
  {http://dx.doi.org/10.1016/S0550-3213(99)00459-9}
  {\path{doi:10.1016/S0550-3213(99)00459-9}}.

\bibitem{Weldon:1982bn}
H.~A. Weldon, {Effective Fermion Masses of Order gT in High Temperature Gauge
  Theories with Exact Chiral Invariance}, Phys. Rev. D26 (1982) 2789.
\newblock \href {http://dx.doi.org/10.1103/PhysRevD.26.2789}
  {\path{doi:10.1103/PhysRevD.26.2789}}.

\bibitem{Burnier:2005hp}
Y.~Burnier, M.~Laine, M.~Shaposhnikov, {Baryon and lepton number violation
  rates across the electroweak crossover}, JCAP 0602 (2006) 007.
\newblock \href {http://arxiv.org/abs/hep-ph/0511246}
  {\path{arXiv:hep-ph/0511246}}, \href
  {http://dx.doi.org/10.1088/1475-7516/2006/02/007}
  {\path{doi:10.1088/1475-7516/2006/02/007}}.

\bibitem{Casas:2001sr}
J.~A. Casas, A.~Ibarra, {Oscillating neutrinos and $\mu \to e\gamma$}, Nucl.
  Phys. B618 (2001) 171--204.
\newblock \href {http://arxiv.org/abs/hep-ph/0103065}
  {\path{arXiv:hep-ph/0103065}}, \href
  {http://dx.doi.org/10.1016/S0550-3213(01)00475-8}
  {\path{doi:10.1016/S0550-3213(01)00475-8}}.

\end{thebibliography}
\end{document}